\documentclass[prb,twocolumn,showpacs,superscriptaddress]{revtex4}
\usepackage{graphicx,bm,amssymb,amsmath}

\newcommand{\Eq}[1]{Eq.~\eqref{#1}}
\newcommand{\eq}[1]{\eqref{#1}}

\newcommand{\beq}{\begin{equation}}
\newcommand{\eeq}{\end{equation}}

\newcommand{\beqa}{\begin{eqnarray}}
\newcommand{\eeqa}{\end{eqnarray}}

\newcommand{\Beqa}{\begin{eqnarray*}}
\newcommand{\Eeqa}{\end{eqnarray*}}

\DeclareMathOperator{\sgn}{sgn}

\DeclareMathOperator{\im}{\mathop{\rm Im}\nolimits}

\newcommand{\rr}{\ensuremath{\mathbf r}}

\newcommand{\pp}{\ensuremath{\mathbf p}}

\newcommand{\PRL}[3]{Phys. Rev. Lett.~\textbf{#1}, #2 (#3)}
\newcommand{\PRB}[3]{Phys. Rev. B~\textbf{#1}, #2 (#3)}

\newcommand{\Science}[3]{Science~\textbf{#1}, #2 (#3)}
\newcommand{\Nature}[3]{Nature~\textbf{#1}, #2 (#3)}
\newcommand{\EPL}[3]{Europhys. Lett.~\textbf{#1}, #2 (#3)}
\newcommand{\JETPLett}[3]{JETP Lett.~\textbf{#1}, #2 (#3)}

\newcommand{\ZhETFLett}[3]{Pis'ma v Zh.E.T.F.~\textbf{#1}, #2 (#3)}

\newcommand{\ZPB}[3]{Zeits. Phys. B ~\textbf{#1}, #2 (#3)}
\newcommand{\JACS}[3]{J. Am. Chem. Soc.~\textbf{#1}, #2 (#3)}

\begin{document}

\title{Permanent polarization of small metallic particles}

\author{A.V. Shytov}
\affiliation{Condensed Matter Physics and Material Science  Department, Brookhaven National Laboratory, 
Upton, NY 11973}
\author{M. Pustilnik}
\affiliation{School of Physics, Georgia Institute of Technology, 
Atlanta, GA 30332}

\begin{abstract}
Electric charge density in a metallic particle 
fluctuates due to inhomogeneities of various kinds. 
While in the bulk of the particle the charge fluctuations are suppressed
by Thomas-Fermi screening,  the underscreened charges near the surface  
give rise to a permanent electric dipole moment.
We show that the dipole moment increases linearly with the
particle size, and fluctuates strongly from particle to particle. 
\end{abstract}

\maketitle

Small metallic particles, even electrically neutral, generate an electric field 
around them. The field arises because the charge density within a particle 
is not uniform. Generally, the charge density includes both a
\textit{dynamic} (time-dependent) contribution, and a \textit{static}, 
one. While the  former is the origin of van der Waals interaction, 
the latter results in a formation of a finite \textit{static} electric dipole 
moment. The static dipoles may determine the interaction between
particles at large distances. Detailed understanding of the interaction 
between nanoparticles is important, e.g., for a self-assembly of
functional nanostructures.\cite{self-assembly} In particular, the dipolar 
interaction is responsible\cite{Murray} for the tendency of nanoparticles 
to arrange in quasi-one-dimensional chains.

The dipole moments of nanoparticles can be measured in 
Stern-Gerlach-type beam deflection experiments,\cite{deHeer,Becker}
or by studying the electric response of dilute solutions of 
nanoparticles.\cite{Alivisatos} Although none of these methods 
produces a reliable set of data for a sufficiently broad range of 
particle sizes, measured values of the static dipole moment 
apparently increase with the particle size. 

While for very small particles (with number of atoms $N\lesssim 10$) 
the formation of the dipole moment can be studied quantitatively 
by \textit{ab initio} methods,\cite{Pickett}, the problem becomes 
untractable for larger~$N$. Indeed, the spatial arrangement of atoms 
may differ considerably from particle to particle. The variations in 
shape and structural defects make the problem even more complex. 

One can view a metallic particle as consisting of a rigid ``skeleton" 
formed by charged ions, and conduction electrons moving freely on this 
background. The skeleton's charge distribution includes a \textit{random} 
component representing structural defects, surface roughness, etc. 
Such structural disorder is present even if the bulk material with the 
same chemical composition is crystalline. It is therefore natural to 
analyze properties of nanoparticles statistically, averaging quantities
of interest over the disorder realizations. In this paper, we apply this 
strategy to the problem of evaluation of the static dipole moment.
We compute the disorder average of $\pp^2$ and find
\begin{equation}
\left\langle \pp^2 \right\rangle \sim e^2 S, 
\label{main-result}
\end{equation}
where $S \sim L^2$ is the surface area of the particle, and $e$ 
is the charge of an electron. (Note that~$\langle \pp \rangle$
 vanishes on symmetry grounds.) We derive \Eq{main-result} 
by considering the disorder-induced mesoscopic density fluctuations. 
We show that the fluctuations originate on a short scale of the order 
of an interatomic distance. The density fluctuations are screened 
in the bulk of the particle. Near the particle's surface, however, 
the fluctuating charges and their screening ``clouds'' form dipoles. 
Adding these random dipoles, we arrive at \Eq{main-result}, which 
implies that the typical value of the dipole moment scales with the 
particle size $L$ as $p \sim \sqrt{\langle p^2\rangle}\sim eL$. As 
discussed towards the end of the paper, the scaling $p\sim eL$ 
is in apparent agreement with the existing experimental data. 

We start the analysis by noting that in a metallic particle with a 
large number of electrons, all electronic states 
below the Fermi level~$\epsilon_F$ contribute 
to the static electron density~$n(\rr)$. 
Quantities of this type can usually be estimated semiclassically. 
Such a consideration yields a homogeneous density~$n(\rr)$ 
that compensates exactly the ion background charge, 
$n(\rr) = n_{\rm ion} = \text{const}$.
However, the semiclassical description breaks down in the vicinity 
of defects or boundaries. Indeed, an impurity embedded in a Fermi 
gas perturbs the density around it. The excess density
$\delta n(\rr) = n(\rr)-n_{\rm ion}$ is described by the Friedel 
oscillation which decays with the distance~$r$ to the impurity as 
$(\lambda_F/r)^3$, and oscillates with the spatial period of the 
Fermi wavelength~$\lambda_F$. (Note that disorder does not destroy 
Friedel oscillations.\cite{Spivak}) Accordingly, the disorder-induced 
density fluctuations are statistically correlated on the scale of $\lambda_F$. 

In order to estimate the short-range contribution to the density 
fluctuations, we consider an \textit{infinite} system with randomly located 
identical pointlike impurities. By the Friedel sum rule, the change of the 
total number of electrons due to a single impurity is given by $\delta_0/\pi$, 
where~$\delta_0$ is $s$-wave scattering phase shift.  To lowest order 
in $|\delta_0|\ll 1$, contributions to $\delta n(\rr)$ from different impurities 
are additive. Neglecting the details of the charge distribution on the scale 
of the order of $\lambda_F$, one can relate the excess density to the 
local impurity concentration, $\delta n (\rr) \approx 2(\delta_0/\pi) n_i (\rr)$, 
with the factor of 2 accounting for the spin degeneracy. Averaging over 
the positions of the impurities then yields
\begin{equation}
K(\rr-\rr') 
= \langle \delta n (\rr) \delta n (\rr') \rangle 
= 4 n_i \left({\delta_0}/\pi\right)^2 \delta (\rr - \rr'),
\label{n-fluct}
\end{equation}
where $n_i = \langle n_{i}(\rr) \rangle$ is the average impurity concentration. 
The phase shift $\delta_0$ can be expressed via the impurity scattering 
cross-section~$\sigma_i =  \lambda_F^2 \delta_0^2\!/\pi$.
Relating the latter to the  elastic mean free path~$l = (n_i\sigma_i)^{-1}$,  
we rewrite \Eq{n-fluct} as 
\begin{equation}
\label{nn-estimate}
K(\rr-\rr')
= 
K_\text{bulk} \delta (\rr - \rr')\ , 
  \qquad
K_\text{bulk} \equiv \frac{4}{\pi}\frac{1}{\lambda_F^2 l}\ .
\end{equation}
Below we show that the estimate \eq{nn-estimate} coincides with the 
result of a rigorous calculation. Before doing so, we compare 
\Eq{nn-estimate} with the results of Refs.~[\onlinecite{BA}] and 
[\onlinecite{Zwerger}]. These authors considered only a contribution 
of electronic states near the Fermi level, treating them in the diffusion 
approximation. Due to a relatively small number of such states, the 
corresponding contribution to~$K(\rr-\rr')$ is smaller by a factor
$(\lambda_F/l)^2\ll 1$ than the estimate \eq{nn-estimate}. For the same 
reason, a small factor~$(\lambda_F / l)^2$ appears in the result of 
Ref.~[\onlinecite{BA}] for the static dipole moment. Thus, \Eq{nn-estimate} 
indeed describes the dominant contribution to the static density fluctuations 
in a disordered metal.
It should be emphasized that due to a large number of electronic
states contributing to density fluctuations, \Eq{nn-estimate} is not 
affected by quantum interference effects such as weak localization.

We now calculate the static density-density correlation function 
microscopically. We consider first noninteracting electrons described 
by the single-particle Hamiltonian
\beq
\hat{H} = -\frac{\nabla^2}{2m} + U (\rr) .
\label{3}
\eeq
Since the density fluctuations are correlated on the scale $\lambda_F\ll L$, 
we ignore the boundaries for the time being and consider an infinite system. 
The random potential~$U(\rr)$ in \Eq{3} represents the structural disorder, 
which is correlated at distances of the order of the interatomic spacing. 
For simplicity, we assume this correlation length to be small compared 
to $\lambda_F$, and write 
\beq
\langle U(\rr) \rangle = 0\ , \qquad
\langle  U(\rr) U (\rr') \rangle 
= \frac{1}{2\pi \nu \tau}\, \delta (\rr - \rr') \ , 
\label{4}
\eeq
where $\nu = m p_F / 2\pi^2$ is the density of states, $\tau$ is
the mean free time, and $p_F$ is the Fermi momentum.
(A more realistic form of \Eq{4} would only affect 
the numerical coefficient in the expression for~$K_{\rm bulk}$.)
Introducing the causal Green's function 
$G(t, \rr, \rr') = -i \langle 
\hat{\mathop{\rm T}\nolimits} \psi^{+}(t, \rr) \psi(\rr')\rangle$, 
and taking into account the spin degeneracy, we write the electron 
density as
$n(\rr) = 2 i G (+0, \rr, \rr)$. 
In frequency domain, this relation takes the form
\beq
n(\rr) = -\frac{2}{\pi}\!\int\limits_{0}^{\epsilon_F}\!d\epsilon\,
\im G(\epsilon, \rr, \rr). 
\label{5}
\eeq

\begin{figure}
\centering
\includegraphics[width=0.99\columnwidth]{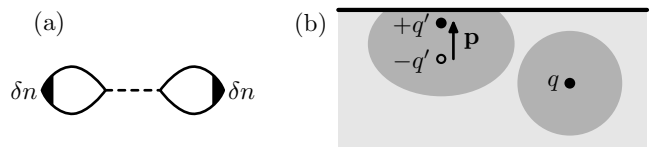}
\caption{(a) The lowest order contribution to the static density-density 
correlation function.
The solid and dashed lines represent electron Green's functions and disorder
potential, respectively. (b) Screening cloud in a metal. A point charge~$q$ in
the bulk induces a symmetric screening cloud.
A charge $q'$ near the surface induces an asymmetric screening cloud with
the screen charge $-q'$ centered at a different point. 
This offset results in a finite 
dipole moment.}
\label{fig-1}
\end{figure}

The lowest order in $\lambda_F / l$ contribution to the product of two 
Green's function corresponds to the diagram\cite{AGD}  
shown in Fig.~\ref{fig-1}(a), which gives
\beq
\label{tilde-k-dirty}
K(\rr-\rr') = \frac{4}{2\pi \nu \tau}
\int\limits_0^{\epsilon_F}\!\!\frac{d\epsilon}{2\pi}\!  
\frac{d\epsilon'}{2\pi}\!  
\int\!d^3 
r^{\prime\prime} 
F_{\epsilon} (\rr - \rr'') F_{\epsilon'} (\rr'' - \rr')
\ . 
\eeq
Here $F_\epsilon (r) = \im G_0^2 (\epsilon, \rr)$ with 
\[
G_{0}(\epsilon,\rr) 
= \frac{m}{2\pi r} e^{i p_\epsilon r \sgn(\epsilon - \epsilon_F)}
\,,
\quad
p_\epsilon = \sqrt{2m\epsilon}
\]
being Green's function in the absence of disorder.
Physically, \Eq{tilde-k-dirty} represents a product of two 
Friedel oscillations induced at points $\rr$ and $\rr'$ by a single 
impurity, averaged over the impurity position $\rr''$.  
Analysis of \Eq{tilde-k-dirty} shows that $K(\rr - \rr')$ 
changes on the scale of the order of $\lambda_F$, which is
the shortest length scale in our problem. At larger distances, 
$K(\rr-\rr')$  can be approximated as 
\beq
K(\rr - \rr') \approx K_\text{bulk} \delta(\rr - \rr')
\ ,
\qquad K_\text{bulk} =  \int\!d^3r\,K(\rr)\ .
\label{dirty-noise1}
\eeq
Substituting \Eq{tilde-k-dirty} into \Eq{dirty-noise1}
and using~$p_F = 2\pi / \lambda_F$ and~$\tau = m l /p_F$, 
we obtain~$K_\text{bulk}$ given by~\Eq{nn-estimate}.
 
Formally, \Eq{dirty-noise1} assumes a constant chemical potential and is valid 
only for an infinite system. In a finite system, however, it is 
the number of electrons that is kept constant. A charge pushed away by 
an impurity is spread over the volume of the particle. 
This can be accounted for by introducing an 
appropriate correction to the chemical potential.
For $L\gg\lambda_F$, this amounts~\cite{BM} to the replacement 
$\delta(\rr - \rr')\to\delta(\rr - \rr') - V^{-1}$ 
in \Eq{dirty-noise1}; here $V\sim L^3$ 
is the particle volume. This yields
\beq
K(\rr) \approx K_\text{bulk} \bigl[\delta(\rr) - {1}/{V}\bigr]
\ , 
\qquad
\int_V d^3r K(\rr)  = 0, 
\label{finite-size}
\eeq
which is consistent with charge conservation.

We turn now to the evaluation of the dipole moment.  
It is  related to the density fluctuation
$\delta n ({\bf r})$ as
\beq
\pp = e\!\int\!d^3 r\,\rr\, \delta n (\rr)\ . 
\label{3a}
\eeq
The random vector~$\pp$ vanishes upon averaging over 
the disorder realizations, $\langle \pp \rangle = 0$. 
Its dispersion $\langle \pp^2 \rangle = \langle p^2 \rangle$, 
however, is finite and is expressed via the density-density correlation function,
\beq
\langle p^2 \rangle = e^2\!\int d^3r \int d^3r' (\rr \cdot \rr') 
\langle \delta n (\rr) \delta n (\rr') \rangle.
\eeq
Using here \Eq{finite-size} for $\langle \delta n (\rr) \delta n (\rr') \rangle$, 
one would find 
\beq
\label{pp-non-interacting}
\langle p^2 \rangle \propto e^2 K_\text{bulk}L^5 ,
\eeq
i.e., $p$ grows with particle size~$L$ as $L^{5/2}$.
However, \Eq{finite-size} is valid 
only in the absence of Coulomb interaction between electrons.
The interaction suppresses charge fluctuations at distances 
larger than the screening length~$l_s$. Thus, 
\Eq{pp-non-interacting} is applicable only for~$L \lesssim l_s$. 
Since~$l_s$ is rather short for most metals, this condition
is very restrictive, and interaction must be taken into account. 

Treating disorder and interaction simultaneously is usually a very 
difficult task. It is greatly simplified when the screening length $l_s$ 
is large compared with the Fermi wavelength $\lambda_F$. 
In the limit~$l_s \gg \lambda_F$, one can analyze the effects of the 
disorder and of the interaction  independently of each other. At short 
distances, $|\rr-\rr'|\ll l_s$, the screening is not effective, and the 
density fluctuations are still described by \Eq{nn-estimate}. The 
fluctuations are ``dressed'' by the screening ``clouds'' at longer distances. 
The dressing can be described in Thomas-Fermi approximation, which 
relates the screening density $\delta n_s(\rr)$ to the electrostatic potential 
$\varphi(\rr)$ as $\delta n_{s} (\rr) = -\kappa e\varphi (\rr)$.
Here $\kappa$ is the compressibility of the electron gas;
disorder-induced fluctuations of $\kappa$ are negligible.\cite{BA,BM} 
The potential $\varphi(\bf r)$ satisfies Poisson's equation
\beq
\label{Laplace}
\nabla^2 \varphi  = - 4\pi e \delta n (\rr),
\eeq
where the total density $\delta n (\rr) = \delta n_0 (\rr) + \delta n_s (\rr)$ 
includes both the bare density $\delta n_0 (\rr)$ described by \Eq{finite-size} 
and the screening contribution $\delta n_s (\rr)$. In general, Thomas-Fermi 
approximation assumes a constant chemical potential. To apply it to a finite 
size system, one has to introduce a correction to the chemical potential to 
ensure charge conservation. Such a modification is not needed in our case, 
since the total charge of the bare density fluctuations vanishes, 
$\int d^3r \,\delta n_0 (\rr) = 0$. Therefore, the total screening charge 
vanishes as well.

Excluding $\varphi(\rr)$ from \Eq{8}, we write $\delta n(\rr)$ as 
\beq
\delta n(\rr) 
= \delta n_0 (\rr) 
- l_s^{-2}\!\int\!d^3 r'Q (\rr, \rr')\,\delta n_0 (\rr')
\label{8}
\eeq
where $l_s = (4\pi e^2\kappa)^{-1/2}$ is the screening length, and 
the function $Q(\rr, \rr')$ satisfies the equation
\beq
\left[-\nabla_{\rr}^2 + l_s^{-2}\right]Q (\rr,\rr') = \delta (\rr-\rr').
\label{9}
\eeq
This equation has to be supplemented with the boundary condition
which corresponds to vanishing of the electric field far away from 
the particle. That is, one has to solve Laplace equation $\nabla^2 Q = 0$
in the exterior of the particle and match the values of $Q$ at the particle 
boundary.

The screening described by \Eq{9} occurs on the scale $l_s$. For a 
large particle, $L \gg l_s$, this scale is small compared to the radius 
of curvature ($\sim L$) of the particle's boundary, which allows us 
to treat the boundary as being locally flat. We choose the coordinate 
axes so that $x > 0$ corresponds to the interior of the particle. It is 
convenient to integrate \Eq{9} along the boundary, which makes the 
electrostatic problem one dimensional. \Eq{9} preserves its form, except 
for $\nabla^2$ being replaced by $\partial^2\!/\partial x^2$. Solution of 
the Laplace equation in the exterior ($x < 0$) reads $Q(x, x') = \text{const}$, 
which gives the boundary condition $\partial Q/\partial x = 0$ at $x = 0$. 
Solution in the interior ($x > 0$) can then be found by the method of images, 
\beq
\int Q(\rr, \rr')\, dy' dz'  
= \,l_s\! \left(e^{- |x - x'|/l_s} + e^{-(x + x')/l_s}\right).
\label{flat-solution}
\eeq
With the help of the relations
\beqa
\int\!\! d^3 r'\, Q (\rr, \rr') 
&=& 
l_s^2, 
\label{Q-integrals}
\\
\int\!\! d^3 r'\, \rr'\,Q (\rr, \rr') 
&=&
\hat{\bm x} \, l_s^{2}  \left(x  + l_s e^{-x/l_s}\right),
\label{Q-inegrals-2}
\eeqa
we see that the total charge 
of the distribution~\eq{8} is zero, while the dipole
moment is finite and is given by
\beq
\delta \pp = e l_s\! 
\int\!\!d^3 r\, \hat{\bm x} \delta n_0(\rr)\, e^{-x/l_s}.
\label{12}
\eeq
Thus, the screening transforms a charge $q$ at distance $x$ 
from the surface into a dipole with the dipole moment 
$\sim q l_s \exp (-x/l_s)$. Qualitatively, this means that 
a charge in the bulk (at $x \gg l_s$) is screened by a spherically 
symmetric cloud. However, if the charge is close to the surface
(at $x\lesssim l_s$), the screening cloud is asymmetric, see
Fig.~\ref{fig-1}(b), giving rise to the dipole moment of the order 
of $q l_s$.

For a smooth boundary of an arbitrary shape the coordinates $x$ 
and $x'$ in Eqs.~\eq{flat-solution}-\eq{Q-inegrals-2} should be 
replaced by the distances from points $\rr$ and $\rr'$ to the boundary, 
and vector $\hat{\bm x}$ by the inward unit normal to the boundary 
at the point closest to~$\rr$. Using \Eq{8}, we write the density-density 
correlation function in the form
\beq
\langle \delta n (\rr) \delta n (\rr') \rangle = 
\int d^3r_1 d^3r_2 \, 
\widetilde{Q}(\rr, \rr_1) K (\rr_1 - \rr_2) \widetilde{Q}(\rr_2, \rr') . 
\label{k-screened}
\eeq
Here 
$\widetilde{Q}(\rr, \rr') = Q (\rr, \rr') - l_s^2 \delta (\rr - \rr')$, 
and $K(\rr_1 - \rr_2)$ given by \Eq{finite-size} describes fluctuations 
of $\delta n_0 (\rr)$. Substituting \Eq{k-screened} into \Eq{3a} 
and using the relations \eq{Q-integrals} and \eq{Q-inegrals-2}, 
we find
\beq
\langle p^2 \rangle 
= e^2 l_s^2 \int d^3 r d^3r'\,K(\rr - \rr') e^{-[x(\rr) + x(\rr')]/l_s},
\label{d-surf-int}
\eeq
where $x(\rr)$ is the distance from point $\rr$ to the surface. 
Taking into account that $L\gg l_s$, we obtain
\beq
\langle p^2 \rangle 
=  \frac{1}{2}\,e^2 l_s^3 K_{\rm bulk} S
= \frac{1}{4\pi}
\frac{e^2 l_s^3}{\lambda_F^2 l}\,S \ , 
\label{d-bulk-screened}
\eeq
where~$S\sim L^2$ is the surface area of the particle, and~$K_{\rm bulk}$ 
given by \Eq{nn-estimate} characterizes the bulk disorder. 

The main contribution to the integral in \Eq{d-surf-int} comes 
from a thin layer of thickness ~$l_s\ll L$ near the surface. 
Therefore, the dipole moment is sensitive to the details of the
surface structure, such as roughness, adsorbed
impurity atoms, etc. The effect of the surface disorder can be also
treated by our method. Indeed, similarly to the bulk impurities, 
the surface defects induce short-range density fluctuations. With the 
screening taken into account, the corresponding contribution to the 
dipole moment has the form of \Eq{d-bulk-screened} with $K_\text{bulk}$ 
replaced by an analogous quantity characterizing the surface disorder. 
Thus, accounting for the surface disorder does not affect the scaling
$\langle p^2\rangle\sim e^2 S$.

The above consideration allows us to find not only the dispersion 
$\langle p^2\rangle$, but the entire distribution function of the dipole moment. 
Indeed, since the bare density fluctuations are correlated only at short distances,
the dipole moment is a sum of a large number $(\sim S l_s /\lambda_F^3\gg 1)$ 
of \textit{independent} random contributions. By the Central Limit Theorem, 
the distribution function of $\pp$ is Gaussian,  
$f(\pp) \propto \exp (-\pp^2\!/2\langle p^2\rangle)$. 
Accordingly, the distribution of the absolute value $p = |\pp|$ 
of the dipole moment is given by
\beq
f(p) \propto p^{\,2} \exp \left(-\frac{p^2}{2\langle p^2\rangle}\right).
\eeq
In principle, the statistics of $p$ can be studied in Stern-Gerlach-type 
experiments, see, e.g.,  Ref.~\onlinecite{deHeer}.

Formally, the analysis given in this paper is valid 
only when~$l_s, l \gg \lambda_F$. In order
to estimate the upper bound to the permanent dipole moment, 
we consider the limiting case $l_s \approx l \approx \lambda_F$,  
and find $ p \sim e\sqrt{S} \sim eL$. 
The dipole moment of this order of magnitude corresponds to the
transferring of one electron across the particle. 
This agrees with the existing experimental data.
In Ref.~[\onlinecite{deHeer}], the dipole moment of niobium (Nb)
clusters with $N\approx 90$ atoms was found
to be of the order of~$0.5$ Debye (D) per atom.  
Assuming the cluster to be a sphere of radius~$a (3/4\pi N)^{1/3}$, 
where~$a \approx 3\,\text{\AA}$ is the lattice constant for Nb, 
we estimate the dipole moment per atom to be
$p/N \approx e\sqrt{S/4\pi}/N \approx 0.5\,\text{D}$. 
According to Ref.~[\onlinecite{Alivisatos}], 
$3-5\,\text{nm}$ - size semiconducting (such as CdSe) nanocrystals
have permanent dipole moments in the range~$50-100\,\text{D}$. 
Even though our theory was developed for metallic clusters 
and cannot be directly applied to semiconductors, the estimate based on
\Eq{d-bulk-screened} gives~$p \sim 100\,\text{D}$. 

Although the static dipole moment $p$ grows linearly with the particle 
size $L$, the electric field induced by the particle in its exterior, 
$E\sim p/L^3\sim e/L^2$, vanishes in the limit $L\to\infty$. Also,
the \textit{polarization} (dipole moment per volume) $p/V$
vanishes in the limit $L\to\infty$, as expected for a normal metal. 
This behavior is to be contrasted with that of ferroelectrics
where $p/V\to \text{const}$ in the thermodynamic limit. 

On the other hand, the main contribution to the disorder-induced density 
fluctuations that give rise to the finite dipole moment comes from a large 
number of the deep-lying electronic energy levels.\cite{remark-2}
Therefore, we expect the static dipole moment to be independent of 
temperature $T$ as long as $T$ is small compared with the Fermi energy. 
However, at $T\gtrsim e^2\!/L$, thermal fluctuations of the dipole moment 
exceed its static value.

To conclude, in this paper we have identified the dominant contribution 
to the charge density fluctuations in a disordered metal and applied
the result to the evaluation of the permanent dipole moment of 
a metallic particle. 

We thank P. Allen, W. de Heer, and U. Landman
for valuable discussions. This work is supported by the DOE 
(contract No. DEAC 02-98 CH 10886), 
and by the NSF (grant No. DMR-0604107).

\end{document}